\documentclass[pre,aps,twocolumn,superscriptaddress,showpacs]{revtex4-1}

\usepackage{dcolumn}
\usepackage{amssymb}
\usepackage{amsmath}
\usepackage{epsf,epsfig}
\usepackage{color}
\usepackage{graphics,psfrag}
\usepackage{graphicx,psfrag}
\usepackage[usenames,dvipsnames]{xcolor}

\usepackage{color}
\definecolor{darkblue}{rgb}{0,0,0.6}
\definecolor{darkred}{rgb}{0.6,0,0}

\usepackage{hyperref}
\hypersetup{
bookmarksopen=true,
pdftitle="Creep in yield stress fluids: a mesoscopic approach",
pdfauthor="C Liu and L Foini and K Martens and JL Barrat", 
pdftoolbar=false, 
pdfstartview={FitH},		
pdfmenubar=true,			
pdfhighlight=/O,			
colorlinks=true,			
urlcolor=darkblue,
citecolor=darkblue,		
linkcolor=MidnightBlue,	
}

\newcommand \be {\begin{equation}}
\newcommand \ee {\end{equation}}
\newcommand \bea {\begin{eqnarray}}
\newcommand \eea {\end{eqnarray}}

\newcommand{\valabs}[1]{\vert #1\vert}

\begin{document}

\title{Mean-field scenario for the athermal creep dynamics of yield stress fluids}

\author{Chen Liu}
\affiliation{Universit\'e Grenoble Alpes, CNRS, LIPhy, 38000 Grenoble, France}

\author{Kirsten Martens}
\affiliation{Universit\'e Grenoble Alpes, CNRS, LIPhy, 38000 Grenoble, France}

\author{Jean-Louis Barrat}
\affiliation{Universit\'e Grenoble Alpes, CNRS, LIPhy, 38000 Grenoble, France}

\date{\today}

\begin{abstract}
We develop a theoretical description based on an existent mean-field model for the transient dynamics prior to steady flow of yielding materials. The mean-field model not only reproduces the experimentally observed non-linear time dependence of the shear-rate response to an external stress, but also allows for the determination of the different physical processes involved in the onset of the re-acceleration phase after the initial slowing down and a distinct fluidization phase. The  fluidization time displays a  power-law dependence on the distance of the applied stress to an age dependent yield stress, which is not universal but strongly dependent on initial conditions.
\end{abstract}

\maketitle

Yield stress fluids (YSF), such as dense emulsions or pastes, display a rich rheological behavior that has attracted considerable attention recently \cite{bonn2009yield, bonn2015yield}. Stationary flow is typically described by a nonlinear flow curve $\sigma(\dot{\gamma})$, where $\sigma$ is the stress and $\dot{\gamma}$ the deformation rate. However, the flow curve is far from accounting for the full complexity of these systems, that involves an interplay between external driving and internal aging, potentially leading to complex thixotropic behavior. In recent years, many experiments and molecular simulations have tried to reveal this complexity using \textit{creep} experiments, in which the flow rate is measured in response to a fixed stress $\sigma$ applied at a given waiting time $t_w$ after sample preparation \cite{struik1978physical,siebenburger2012creep, baldewa2012delayed,leocmach2014creep,ballesta2016creep,chaudhuri2013onset,sentjabrskaja2015creep,landrum2016delayed}. These experiments, that lead, for  $\sigma$ larger than a yield stress $\sigma_Y$, to flow or failure, reveal an intriguing behavior, with two salient features: (i) the strain-rate $\dot{\gamma}(t)$ in response to a stress larger than the yield stress is strongly non-linear and nonmonotonous, with a so called  "s-shaped" dependence  of $\dot{\gamma}(t)$ \cite{divoux2011stress,siebenburger2012creep,coussot2006aging}, including a nontrivial "primary creep regime"  often described by a power law $t^{-\mu}$.  (ii) The fluidization time scale $\tau_f$ diverges when approaching  yield stress, however in a non-universal manner.

In this work, we develop an approach that explains these features in athermal systems, in which thermal fluctuations have essentially no influence on the flow. These  systems are a large subset of YSF \cite{bonn2015yield}, including e.g. foams, emulsions, physical gels, or granular media. In spite of the irrelevance of thermal fluctuations, the creep dynamics  will depend on the initial condition determined by the preparation process and  the subsequent waiting period, during which slow processes such as coarsening or compaction can alter the level of relaxation. Our approach contrasts previous attempts addressing systems in which thermal fluctuations are important,  based on the  soft glassy rheology model \cite{petersgr1998, fielding2000aging, moorcroft2013criteria,merabia2016creep} or mode-coupling theory \cite{mctreho2002, brader2009glass}, 

Our description is based on a mean-field version of the elasto-plastic scenario, that describes the flow as resulting from interactions among local plastic events triggered by the external driving, and  accounts for the flow properties of athermal YSF  \cite{Argon79, PicardPRE05, NicolasPRL13, PuosiPRE14, LiuPRL16, albaret2016mapping, boioli2017shear}. It extends a previous formulation for imposed shear-rate  \cite{HL} to a system subjected to an imposed stress, allowing us to address complex protocols, in particular  typical creep experiments. Using this approach, we reproduce  features (i) and (ii) above. We also find that the fluidization time is sensitive to  initial aging, represented by an initial distribution of quenched local stresses. The divergence of the  fluidization time is described by a power-law relation if the distance to yield is characterized using an age-dependent static yield stress $\sigma^S_y$. However the exponent appears, as in experiments, to be non-universal.

To investigate the creep dynamics we develop an extension of the H\'ebraud-Lequeux model \cite{HL} which belongs to the class of athermal, local yield stress models \cite{HL,KEP1,agoritsasEPJE15, EliArXiv16}. The distribution $\mathcal{P}(\sigma,t)$ characterize the stress values $\sigma$ in a subvolume of mesoscopic size. The macroscopic stress, which in creep is the externally applied load,  $\sigma^\mathrm{ext}$ is computed as $\sigma^\mathrm{ext}\equiv\langle\sigma\rangle(t)\hat{=}\int\mathcal{P}(\sigma,t)\sigma d\sigma$. 

The evolution equation for $\mathcal{P}$ is formulated in terms of a time dependent strain-rate,  $\dot\gamma(t)$, as:
\begin{eqnarray}
 \partial_t \mathcal{P}(\sigma,t) &=& - G_0 \dot{\gamma}(t)  \partial_\sigma \mathcal{P}(\sigma,t) - \frac{1}{\tau}\theta(\valabs{\sigma}-\sigma_c)\mathcal{P}(\sigma,t) \nonumber\\
 && + \Gamma(t) \delta(\sigma) + D(t) \partial_\sigma^2 \mathcal{P}(\sigma,t)\;.
\label{eq:hl}
\end{eqnarray}
The first term on the right hand side accounts for the local elastic response with shear modulus $G_0$, the second term describes local yielding at rate $1/\tau$ if the local stress exceeds a threshold $\sigma_c$ ($\theta$ is the  Heaviside distribution) and the third term is the  gain term accounting for a complete relaxation of the stress, where $\delta(\sigma)$ is the Dirac distribution. The last term is a mean-field description of the interaction generated by stress redistribution after local yielding. It describes the resulting  fluctuations as a diffusive  process with a time-dependent diffusion constant $D(t)$ proportional to the rate of plasticity $\Gamma(t)$: 
\begin{eqnarray}
D(t) = \alpha \Gamma(t) \;\;\; \mathrm{ and } \;\;\; \Gamma(t)= \frac{1}{\tau} \int_{\valabs{\sigma}>\sigma_c} d\sigma \mathcal{P}(\sigma,t) 
\label{eq:hl2}
\end{eqnarray}
The mechanical coupling strength $\alpha$ characterizes how strongly local stresses are altered by surrounding plastic events \cite{PuosiSoftMatter2015}. For a fixed shear-rate and $\alpha < 1/2$ the model leads to a yield stress rheology, admitting a finite value of the average stress $\sigma_y(\alpha) < \sigma_c$ (dynamical yield stress) for vanishing strain-rate.

To implement a controlled stress protocol, the evolution of the shear-rate is constrained to follow the plastic shear-rate according to 
\begin{eqnarray}
\dot{\gamma}(t) = \frac{1}{\tau G_0}\int_{\valabs{\sigma}>\sigma_c} d\sigma \sigma  \mathcal{P}(\sigma,t) 
\label{eq:gammadot}
\end{eqnarray}
Using (1), it is easily verified that the average stress remains constant during the evolution of $\mathcal{P}(\sigma,t)$. 
The desired value $\sigma^\mathrm{ext}$ is imposed through the initial stress distribution $\mathcal{P}(\sigma,t=0)$.  
 
The initial condition is defined by the  stress distribution $\mathcal{P}_0(\sigma)= \mathcal{P}(\sigma,t=0)$. To mimic situations of a quenched system before applying the stress step, we consider distributions with zero mean $\mathcal{P}^I_0(\sigma^\mathrm{int})$, instantaneously shifted by the desired value of the applied stress $\sigma^\mathrm{ext}$ at the onset of the experiment, i.e $\mathcal{P}_0(\sigma)=\mathcal{P}^I_0(\sigma-\sigma^\mathrm{ext})$. In principle we should consider only distributions with a compact support $\sigma<\sigma_c$, so that the system does not evolve until the external load is applied. Hence, the model does not display aging.

In a first approach, we assume for $\mathcal{P}^I_0(\sigma^\mathrm{int})$ a Gaussian shape \footnote{Strictly speaking this initial condition violates the  condition of a compact support, but in practice the standard deviations studied are small enough such that the statistical weight beyond $\sigma_c$ can be regarded negligible.} centered at zero \cite{srolovitz1981radial}. The only parameter is the standard deviation $s_d$, characterizing the level of residual heterogeneity in an amorphous system. As more relaxed systems display a less prominent Boson peak, which is indicative of a better homogeneity of the elastic properties \cite{duval2006physical}, we assume relaxation is also reducing the width of the stress distribution. Thus a $\mathcal{P}^I_0(\sigma^\mathrm{int})$ with a smaller $s_d$ corresponds to a more relaxed system, and we will take $s_d$ as an indirect measure of the age. Interestingly the standard deviation of our distribution can be formally linked to the aging parameter in the lambda-model for thixotropic materials \cite{wei2016quantitative,armstrong2016dynamic}, as discussed in the supplementary material.

We numerically solve Eq.(\ref{eq:hl}) using an explicit Euler method. The dynamic yield stress $\sigma_y$ of the mean-field model is a decreasing function of the mechanical coupling strength $\alpha$ \cite{agoritsasEPJE15, PuosiSoftMatter2015}. Our range of investigation is restricted to $s_d \geq 0.22$ and $\sigma^{ext} -\sigma_y(\alpha)\geq 0.12$ due to numerical limitations.

\begin{figure}[th]
\begin{center}
\includegraphics[width=\columnwidth, clip]{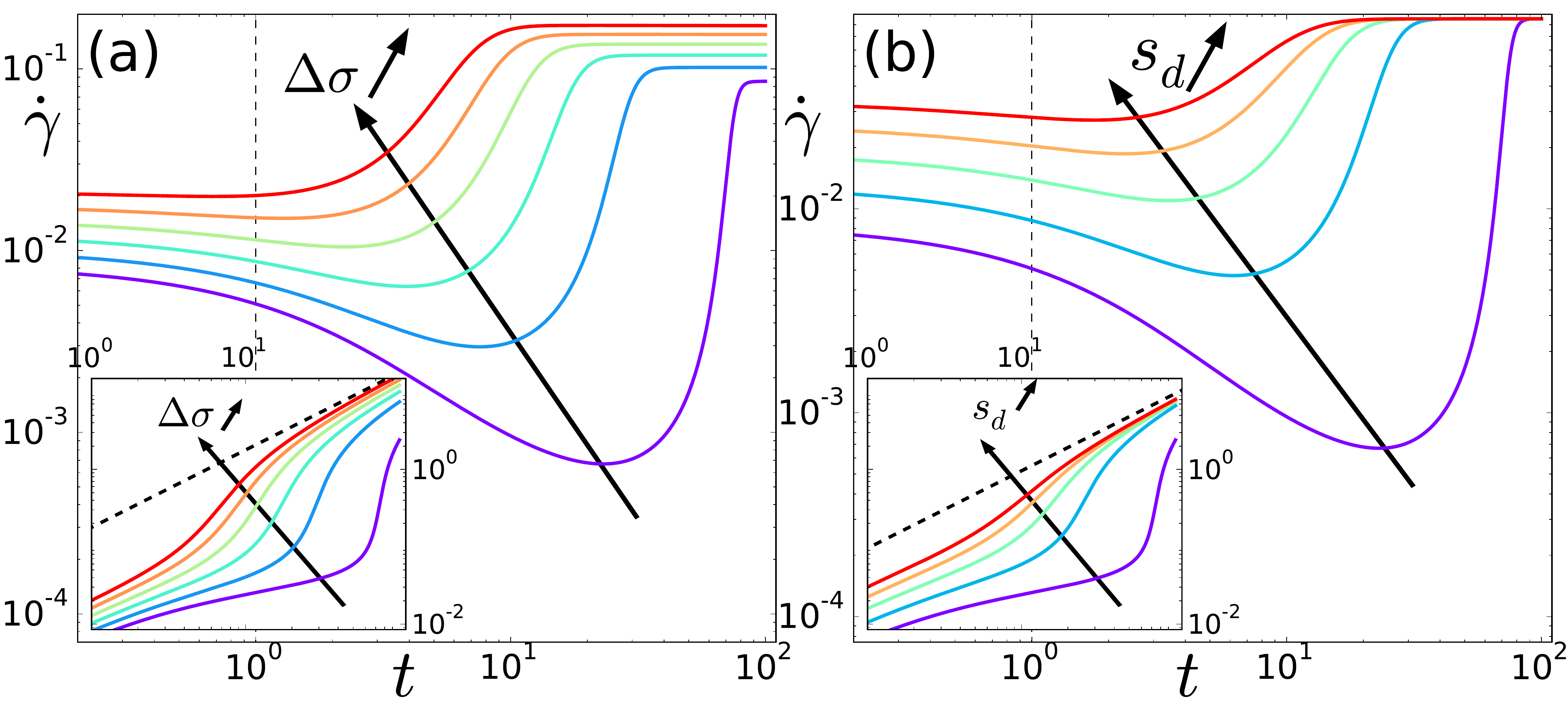}
\end{center}
\caption{{\it Creep curves} Strain-rate time series $\dot{\gamma}(t)$ and corresponding strain time series $\gamma(t)$ (inset), dashed linear line serves as a guide for the eye. (a) Creep behavior for different applied stresses for a fixed initial aging level $s_d=0.28$ with $\Delta\sigma=0.18$, $0.2$, ..., $0.28$. (b) Creep behavior for a fixed applied stress $\Delta\sigma=0.18$ for different initial aging levels $s_d=0.28$, $0.3$, ..., $0.36$.}
\label{fig:creep_mf}
\end{figure}

{\it Results -- } Firstly, a typical non-linear response of $\dot{\gamma}(t)$ and $\gamma(t)$ after application of a step stress $\sigma^{ext}$ larger than $\sigma_y$ is shown in Fig.~\ref{fig:creep_mf}. At $t\approx1$, $\dot{\gamma}(t)$ begins to evolve significantly. For small enough $\sigma^{ext}$, the mean-field model reproduces the characteristic s-shaped curve for $\dot{\gamma}(t)$ that has been observed in various experiments \cite{coussot2006aging,siebenburger2012creep,divoux2011stress}. For a fixed initial aging level ($s_d$ fixed in Fig.~\ref{fig:creep_mf}(a)), before entering the  plateau of the steady-state regime, $\dot{\gamma}(t)$ displays a creep regime where the shear-rate decreases with time in an apparent power-law until it reaches a minimum. Within this creep regime, the accumulated strain $\gamma(t)$  shows a sub-linear increase in time. After the minimum in $\dot{\gamma}$, the system enters a fluidization regime where the shear-rate speeds up toward the steady-state, and correspondingly the accumulated strain $\gamma(t)$ increases super-linearly to reach the  linear regime of a fluid. As the applied stress increases, the extent of the creep regime decreases,  until it eventually disappears and the system enters directly the fluidization regime leading to steady-state flow (Fig.~\ref{fig:creep_mf}(a)). A similar effect of the initial aging level is found for a given applied stress( Fig.~\ref{fig:creep_mf}(b)). The duration of the creep regime decreases when increasing $s_d$ (decreasing age), until for a large enough  $s_d$  the creep regime disappears and finally reaches the same stationary shear-rate. This type of dependence on the applied stress and initial aging is similar to what has been reported for bentonite suspensions \cite{coussot2006aging} and colloidal hard spheres \cite{siebenburger2012creep}.

Several works in the literature describe the slowing down as a power-law  $\dot{\gamma}\sim t^{-\mu}$, expecting that the creep exponent $\mu$ may have universal features. Our results can indeed be fitted using such a power law, in particular when the applied stress $\sigma^{ext}$ gets small. However, we observe that the exponent $\mu$ of the apparent power-law decreases with increasing applied stress and decreasing initial aging (increasing $s_d$). We measure, for those $\dot{\gamma}(t)$ exhibiting a creep regime (defined as from $t=1$ to the minimum of $\dot{\gamma}$) that lasts at least one decade, the exponent $\mu$ varying from $\approx 1.2$ to $\approx 0.6$, a range comparable to the one reported in  experimental studies \cite{divoux2011stress,ballesta2016creep,landrum2016delayed}.

We now discuss the relation between the fluidization time scale and the distance of the applied stress to the dynamical yield stress $\Delta\sigma =\sigma^\mathrm{ext}-\sigma _y$. Two time scales can be identified from the time dependence of the shear rate:  $\tau _m$, corresponding to the  minimum of $\dot{\gamma}$ and $\tau_f$ defined as the inflection point of  $\dot{\gamma}(t)$ before entering the steady flow region.  Following \cite{divoux2011stress}, we identify $\tau_f$ as the fluidization time, which can be defined even in the absence of a well developed creep regime. 
We will show that $\tau_f$ and $\tau _m$ are associated with different mechanisms, and $\tau_f$ is actually the adequate choice to characterize the time scale for entering the stationary flow.

\begin{figure}[th]
\begin{center}
\includegraphics[width=\columnwidth, clip]{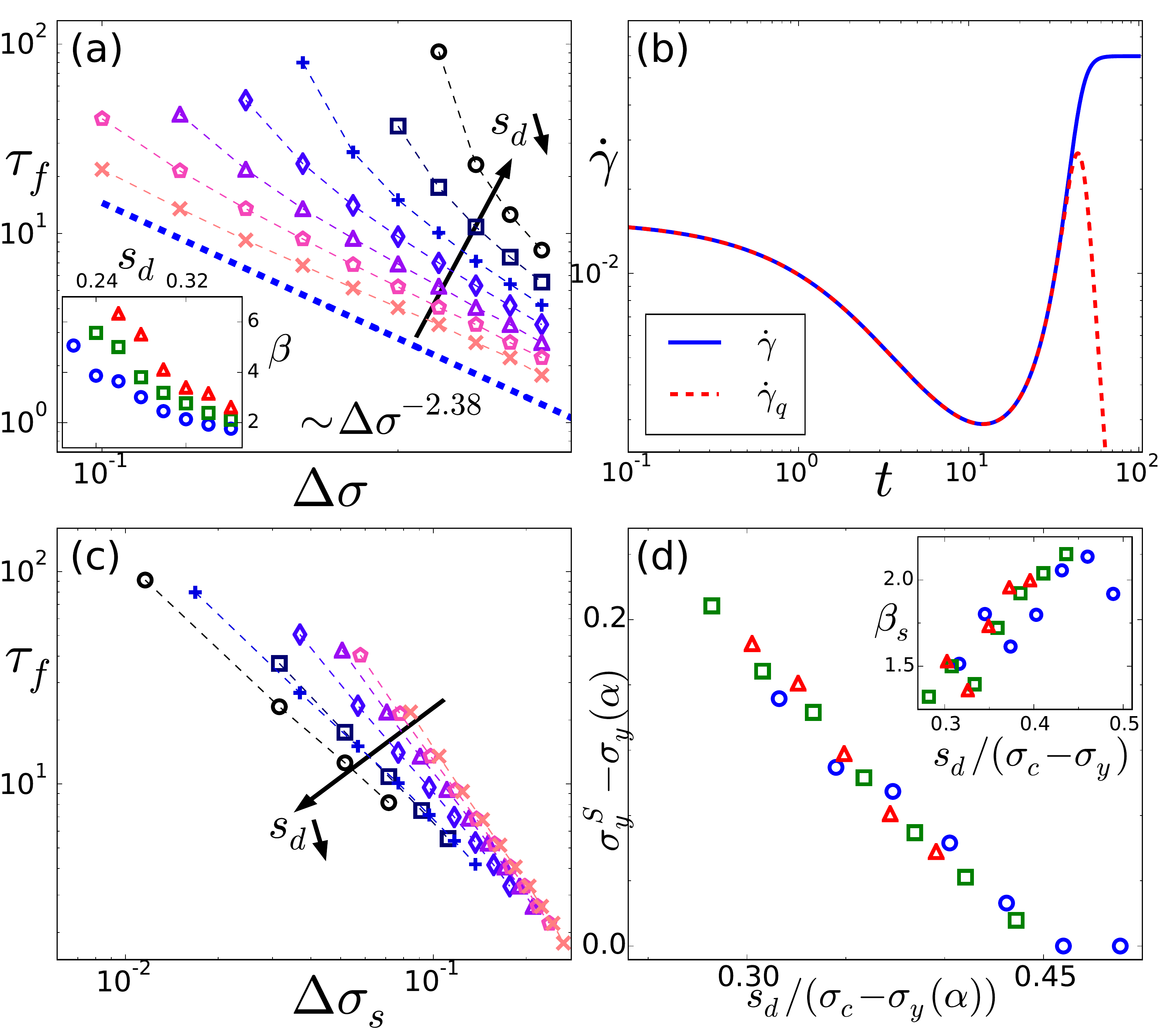}
\end{center}
\caption{{\it Creep behavior.} (a): Fluidization time $\tau_f$ versus $\Delta\sigma=\sigma-\sigma_y$ for $\alpha=0.3$ and $s_d = 0.22, 0.24, ...,0.34$. Inset: $\beta$ v.s. $s_d$ for $\alpha=0.4$ (triangle), $0.3$ (square), $0.2$ (circle). (b): Comparison between $\dot{\gamma}_q(t)$ (dashed curve) generated by Eq.(\ref{eq:hl_pq}) and $\dot{\gamma}(t)$ (solid curve) generated by Eq.(\ref{eq:hl}). (c): $\tau_f$ v.s. $\Delta\sigma_s$ for $\alpha=0.3$. Different symbols for different values of $s_d$, the correspondence is the same as in (a). (d): The difference between static and dynamic yield stress $\sigma^S_y-\sigma_y$ against the normalized initial relaxation $s_d/(\sigma_c-\sigma_y)$. Inset: $\beta_s$ v.s. $s_d/(\sigma_c-\sigma_y)$. $\alpha=0.4$ (triangle), $0.3$ (square), $0.2$ (circle).}
\label{fig:interpts}
\end{figure}

Typical relations between $\tau_f$ and $\Delta\sigma$ for different aging levels are shown in Fig.~\ref{fig:interpts}(a). Previous experimental results \cite{divoux2011stress} suggest $\tau_f\sim \Delta\sigma ^{-\beta}$ with $\beta$ measured from $2$ to $8$ depending on the sample preparation. Note that this behavior must be distinguished from studies on thermal systems \cite{merabia2016creep,lindstrom2012structures,sprakel2011stress,gibaud2009shear} which suggest an exponential relation. Here our results show convexity for $s_d$ small (i.e. well relaxed systems) (Fig\ref{fig:interpts}(a)) indicating the fluidization time increases faster than a power-law as $\Delta\sigma$ approaches zero. For larger  $s_d$,  $\tau _f(\Delta\sigma)$ becomes closer to a power-law. The longer the waiting time before the startup of the creep protocol, the stronger the fluidization time increases for decreasing $\Delta\sigma$ (Fig.~\ref{fig:interpts}(a)).

To quantify the dependence of $\tau _f(\Delta\sigma)$ on $s_d$, we fit each curve in Fig \ref{fig:interpts}(a) with a power law $\tau _f = A \Delta\sigma^{-\beta}$. As discussed above this fitting form is appropriate only for large enough $s_d$. For small $s_d$, the reported value is an effective exponent $\beta = \max \frac{d\ln \tau_f}{d\Delta \sigma} $, and carries qualitative information on how fast the fluidization process slows down with decreasing stress. The result is shown in the inset of Fig\ref{fig:interpts}(a). A systematic decrease of $\beta$ with  $s_d$ is observed for all $\alpha$. Note that the $\beta$ value measured for carbopol microgel \cite{divoux2011stress,grenard2014timescales} lies in the same range (from $2$ to $8$) as found here. We also notice that the exponent $\beta$ depends on  the strength $\alpha$ of the  mechanical coupling. A larger $\alpha$ yields a larger $\beta$ for the same $s_d$.
 
An analytic study of the complex transient dynamics is difficult, due to the strong non-linearity of Eq.~(\ref{eq:hl}). Therefore, we limit our analysis to a semi-qualitative discussion. Initially, the population of marginal stable nodes which is the source of plastic activity, decreases exponentially in time (the term $\frac{1}{\tau}\Theta(\big|\sigma\big|-\sigma _c)\mathcal{P}$) but is also enhanced by the population of nodes having a stress slightly smaller than $\sigma_c$. This enhancement results from two comparable sources: a drift and a diffusion corresponding respectively to the first and second partial derivative term in Eq.(\ref{eq:hl}). Note that both contributions are proportional to the  population of marginally stable nodes. Further the drift and diffusion induced fluxes are, respectively, proportional to $\mathcal{P}\big|_{\sigma = \sigma_c}$ and $\partial_{\sigma}\mathcal{P}\big|_{\sigma=\sigma_c}$.

Let us first consider two extreme cases with $\sigma^\mathrm{ext}>\sigma_y$, where a steadily flowing state exists according to the flow curve. If the standard deviation of the initial  distribution $\mathcal{P}_0(\sigma)$ is large enough, the supply of unstable sites compensates the losses, and the flow will  be strongly accelerated until  steady state. This  corresponds to the transient curves without any creep regime in Fig.~\ref{fig:creep_mf}. In contrast, if the  initial Gaussian distribution has a small standard deviation, not only a small portion of the population is marginally stable, but also the values of $\mathcal{P}\big|_{\sigma = \sigma_c}$ and $\partial_{\sigma}\mathcal{P}\big|_{\sigma=\sigma_c}$ are close to zero. As a consequence the drift and the diffusion terms are weak and become even weaker since the marginally stable population decreases exponentially. The shear rate decreases rapidly, and the system gets stuck in a configuration where all stresses are below $\sigma_c$ and the flow stops. Note that this situation can be observed in experiments and simulations \cite{bonn2015yield} even if the applied external stress is larger than the dynamic yield stress $\sigma_y(\alpha)$.

The above analysis raises the issue of the evolution of $\mathcal{P}(\sigma,t)$ that causes the transition from the creep regime to the fluidization regime and eventually the steady flow for intermediate values of $s_d$. It suggests further that $\tau_f$ diverges \textit{before} $\Delta\sigma$ tends to zero (see Fig.~\ref{fig:interpts}(a)). For a  well relaxed system (small $s_d$), the static yield stress $\sigma^S_y$, which is the minimum stress needed to fluidize a system at rest, is larger than the dynamical one  $\sigma_y$. In an extreme situation where $s_d=0$, one should apply $\sigma^\mathrm{ext}\geq\sigma^S_y=\sigma_c$ to make the system flow. This is consistent with previous studies on transient dynamics  that report on an age dependent overshoot in the stress-strain curve \cite{fielding2000aging,rottlerrobbins,rodney2011potential}. A comparison between $\sigma^S_y$ and the stress overshoot is presented in our supplementary material.

To gain a better understanding of the initial evolution of the shear-rate, we now approximate the full dynamics in the early regime, where mesoscopic subvolumes have been activated at most once, by setting $\mathcal{P}=\mathcal{P}_q+\mathcal{P}_a$ with $\mathcal{P}_q$ referring to the sites that have never been activated,  and $\mathcal{P}_a$  to those activated once. Thus  $\mathcal{P}_q(\sigma,t=0)=\mathcal{P}_0(\sigma)$, $\mathcal{P}_a(\sigma,t=0)=0$ and the distributions obey:
\begin{eqnarray}\label{hl_pq}
\partial_t \mathcal{P}_q(\sigma,t) &=& - G_0 \dot{\gamma}_q(t)  \partial_\sigma \mathcal{P}_q(\sigma,t) + D_q(t) \partial_\sigma^2 \mathcal{P}_q(\sigma,t) \nonumber\\
 && - \frac{1}{\tau}\theta(\valabs{\sigma}-\sigma_c)\mathcal{P}_q(\sigma,t) \;.
\label{eq:hl_pq}
\end{eqnarray}
\begin{eqnarray}\label{hl_pa}
\partial_t \mathcal{P}_a(\sigma,t) &=& - G_0 \dot{\gamma}_q(t)  \partial_\sigma \mathcal{P}_a(\sigma,t) + D_q(t) \partial_\sigma^2 \mathcal{P}_a(\sigma,t) \nonumber\\
 && +\Gamma_q(t)\delta(\sigma) \;.
\label{eq:hl_pa}
\end{eqnarray}
where $\dot{\gamma}_q(t)$, $\Gamma_q(t)$ and $D_q$ are defined as above,  with $\mathcal{P}$ replaced by $\mathcal{P}_q$. We note that (\ref{eq:hl_pq}) and (\ref{eq:hl_pa}) approximate the full dynamics, ignoring the possibility of multiple activation. As a result they will always lead to a vanishing strain rate  at long times, $\dot{\gamma}_q(t)\rightarrow 0$. However, the comparison between this approximation and the full solution will give us insights into the time range over which the initial condition influences the fluidization process.

The approximate solution $\dot{\gamma}_q(t)$ (obtained by solving  (\ref{eq:hl_pq}) and (\ref{eq:hl_pa})) and the full solution $\dot{\gamma}(t)$, for the same initial setting $(\sigma^\mathrm{ext}, s_d)$,  are compared  in Fig.~\ref{fig:interpts}(b). Up to the mid-fluidization, $\dot{\gamma}$ and $\dot{\gamma}_q$ are in good agreement, indicating that the transient dynamics is dominated by sites that undergo their first activation. Within this first stage, the flux of sites transfered from $\mathcal{P}_q$ to $\mathcal{P}_a$ first decreases with time as the population of initially unstable sites gets depleted (corresponding to the creep regime). At $\tau_m$, this flux starts increasing, and $\mathcal{P}_a$ starts to represent a significant fraction of the sites. Beyond $\tau_f$, $\mathcal{P}_a$ becomes dominant and further activation events become essential. Eventually the memory of the initial condition is lost and the steady flow regime can be achieved in the full dynamics of $\mathcal{P}$.

This analysis suggests that the critical yield stress at which fluidization takes place is not determined by the steady-state flow curve, but depends strongly on initial conditions and relaxation level. Hence, a power-law behavior cannot be expected if the dynamic yield stress is taken as reference. Rather, we estimate the static yield stress from the divergence of the fluidization time, by identifying for a given $s_d$ the value for which a power-law $\tau_f \sim \Delta\sigma_s^{-\beta_s}\equiv [\sigma^\mathrm{ext}-\sigma^S_y(s_d)]^{-\beta_s}$ holds. Finding the best power-law fitting we estimate both $\beta_s(s_d)$ and $\sigma^S_y(s_d)$. The result of this analysis is shown in Fig.~\ref{fig:interpts}(c), where a power-law extends to at least one decade. 

The static yield stress and exponent $\beta_s$ as a function of the initial aging  for different couplings $\alpha$ are shown in Fig.~\ref{fig:interpts}(d). We use $s_d/(\sigma_c-\sigma_y)$ as the control variable, since the effects of $s_d$ for different values of $\alpha$ are comparable only when measured relative to the distance between the local threshold $\sigma_c$ and the dynamical yield stress $\sigma_y(\alpha)$. When $s_d$ becomes of the order of $\sigma_c - \sigma _y(\alpha)$, the initial configuration contains a sufficient number of marginally stable nodes, and no difference is expected between static and dynamic yield stress.  
This is confirmed by the data shown in Fig.~\ref{fig:interpts}(d) where $\sigma^S_y-\sigma_y$ decreases to zero as $s_d/(\sigma_c-\sigma_y)$ become of order  $1$. In the opposite limit $s_d=0$, the static yield stress should be equal to $\sigma_c$, as discussed above.  Interestingly  Fig.~\ref{fig:interpts}(d) shows that  $\sigma^S_y-\sigma_y$ as a function of $s_d/(\sigma_c-\sigma_y)$ is described by a master curve independent of  $\alpha$, suggesting a universal normalized relaxation level, at which the static and dynamical yield stress become identical for different systems (characterized by different $\alpha$). The insets of  Fig.~\ref{fig:interpts}(d) shows the exponent $\beta_s$ against $s_d/(\sigma_c-\sigma_y)$.  A clear tendency of increasing in $\beta_s$ with $s_d/(\sigma_c-\sigma_y)$ can be observed until the point where $\sigma^S_y$ becomes identical to $\sigma_y$. In spite of uncertainties associated with the fitting procedure, the collapse of $\beta_s$ for different $\alpha$  also suggests  a master relation between $\beta_s$ and $s_d/(\sigma_c-\sigma_y(\alpha))$. The value of $\beta_s$ at large $s_d$ is comparable with experimental measurements \cite{divoux2011stress,grenard2014timescales}.
 
To summarize, we proposed a mean-field approach based on the H\'ebraud-Lequeux model that allows one to study various rheological protocols in athermal YSF. Using this approach , we study the creep behavior (flow induced by an imposed stress) of YSF starting from different initial levels of relaxation. We reproduce the s-shaped response of the shear-rate after imposing a step stress, in qualitative agreement with experiments \cite{coussot2006aging, siebenburger2012creep}. To be able to compare with experimental results, we quantify the slowing down in the creep regime with a power-law $\dot{\gamma}\sim t^{-\mu}$ and find that the exponent $\mu$ lies in the same range as reported in experiments \cite{divoux2011stress}. The apparent exponent is non-universal, and depends on the applied stress and on the initial relaxation level. We distinguished the different underlying mechanisms for the two time scales $\tau_m$ (minimum strain-rate)  and $\tau_f$ (fluidization). $\tau_m$ is determined by the first plastic activations and depends sensitively on  the initial distribution of internal stresses $\mathcal{P}_0(\sigma)$. $\tau_f$ characterizes the loss of memory with respect to this initial distribution. We rationalized our results for the fluidization time $\tau_f$ by introducing a static yield stress $\sigma^S_y$ that increases with initial relaxation. The behavior of  $\tau_f$ can then be described by  a power law $\tau _f\sim(\sigma^\mathrm{ext}-\sigma^S_y)^{-\beta_s}$, where $\beta_s$ increases with decreasing initial aging, and has again values comparable to those reported in experiments. Finally, we propose that different systems can be compared by introducing an appropriate measure of relaxation based on the distribution of internal stresses, a proposal that could be tested in microscopic simulations and experiments giving access to local stresses.

{\it Acknowledgements -- }
J.-L.B. and C.L. acknowledge financial support from ERC grant ADG20110209.J-L. B is a member of Institut Universitaire de France.
K.M. acknowledges financial support of the French Agence Nationale de la Recherche (ANR), under grant ANR-14-CE32-0005 (FAPRES) and CEFIPRA grant. We thank Laura Foini for detailed discussions and for her help in developing the numerical code.

\newpage

\section*{Supplemental Material for:\\
	``Mean-field scenario for the athermal creep dynamics of yield-stress fluids''
}

	
	In this supplementary material we address three complementary aspects for the main article: First we explain how the static yield stress $\sigma^S_y(\alpha,s_d)$ introduced in the manuscript is related with the stress overshoot in the stress-strain curve obtained by shear start up test in the zero shear rate limit. Second we show that an effective "$\lambda$-model" formulation can be derived from our model. Finally we describe in detail the numerical implementation of our model used and some related numerical limits.
	


\section{Static yield stress and the stress overshoot}

Stress overshoots in shear start up experiments are commonly observed in both dense hard sphere systems \cite{rottlerrobbins}, but also in network forming colloidal gels \cite{jamalie_prl}. In the latter case, the stress overshoot is closely related with the change of structural quantities, while in the former case, which is of our interests, the stress overshoot depends on the relaxation level (age) of the system \cite{fielding2000aging,rottlerrobbins,rodney2011potential}. 

The static yield stress $\sigma^S_y(s_d, \alpha)$ is, according to our study, closely related to the initial relaxation level characterized by $s_d$. In this section we compare $\sigma^S_y$ with the stress overshoot during a shear start up produced by our model. 

\begin{figure}[th]
	\begin{center}
		\includegraphics[width=\columnwidth, clip]{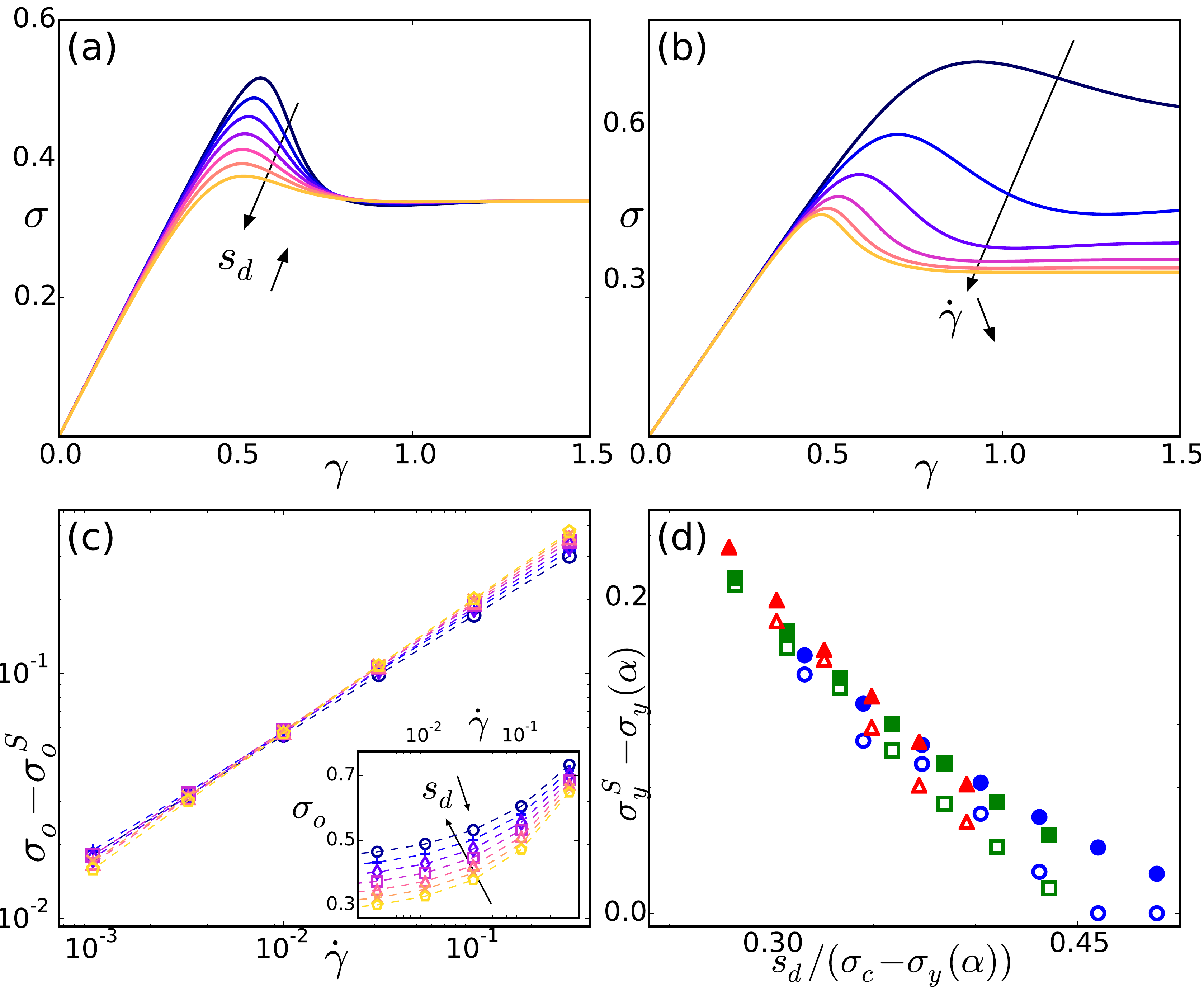}
	\end{center}
	\caption{{}(a) Stress-strain curves for $\alpha=0.2$, $\dot{\gamma}=10^{-2}$, from the top to the bottom $s_d=0.22, 0.24, ...,0.34$. (b) Stress-strain curves for $\alpha=0.2$, $s_d=0.26$, from the top to the bottom $\dot{\gamma}=10^{-0.5}, 10^{-1}, 10^{-1.5},..., 10^{-3}$. (c) The power law dependence between $\sigma_o-\sigma^S_o$ and $\dot{\gamma}$ where $\sigma^S_o$ represents the overshoot stress at zero shear rate limit. Different symbols for different values of $s_d$ for the initial condition. Inset: the corresponding raw data of $\sigma_o$ v.s. $\dot{\gamma}$. (d) Empty symbols and Filled symbols represent respectively $\sigma^S_y-\sigma_y(\alpha)$ and $\sigma^S_o-\sigma_y(\alpha)$ as  function of the normalized initial relaxation level $s_d/\big(\sigma_c-\sigma_y(\alpha)\big)$. Colors for different values of $\alpha$, $\alpha=0.2$(blue), $\alpha=0.3$(green), $\alpha=0.4$(red).}
	\label{fig:2}
\end{figure}

To assess the effect of initial relaxation on the stress overshoot, we perform shear start-up simulations (at constant shear rate) from an initial condition $\mathcal{P}_0(\sigma)$ corresponding to a centered Gaussian with different $s_d$, as used in the main article. The typical stress-strain curves for different $s_d$ at fixed $\dot{\gamma}$ and for different $\dot{\gamma}$ at a fixed $s_d$ are shown, respectively, in Fig.~\ref{fig:2}(a) and Fig.~\ref{fig:2}(b). The overshoot stress $\sigma _o$ is defined as the maximum stress of the stress-strain curve during the shear start-up. These two figures illustrate that the overshoot stress reflects the combined effects of rejuvenation by shearing and initial aging. Since the static yield stress $\sigma^S_y$ is the threshold of stress that should be applied for fluidizing a system at rest with certain initial relaxation level, one should expect that, when the shear rate becomes so slow that the rejuvenation effect becomes negligible, the overshoot stress reflecting only the initial aging should be comparable with the static yield stress. 

The dependence of the overshoot stress $\sigma_o$ on the shear rate $\dot{\gamma}$ for a given $(\alpha, s_d)$ can actually be fitted with a Herschel-Bulkley type relation, i.e. $\sigma_o=\sigma^S_o+B\dot{\gamma}^n$. By adjusting the value of $\sigma^S_o$, a clear linear relation can be found when $\sigma_o-\sigma^S_o$ is plotted against $\dot{\gamma}$ in log-log scale, Fig.~\ref{fig:2}(c).  This allows us to estimate the overshoot stress in the  limit of zero shear rate $\sigma^S_o$ for the given $(\alpha,s_d)$.

In figure \ref{fig:2}(d) we plot  on top of figure(2.d) of the main article, the resulting values of  $\sigma^S_o-\sigma_y$ against $s_d/(\sigma_c-\sigma_y)$ for the same different values of $\alpha$.  For small values of $s_d$, $\sigma^S_y$ agrees well with $\sigma^S_o$, which justifies our idea that the underlying physics of the static yield stress observed in creep experiments and that of the zero shear rate limit stress overshoot in shear start-up experiments are the same. 

We also notice from the figure that for larger values of $s_d$, a small but systematic deviation exists  between $\sigma^S_y$ and $\sigma^S_o$. Such initial conditions correspond to poorly relaxed systems, with a very short fluidisation time, for which deviations from the general scenario outlined in this work may be expected. In any case, a Gaussian distribution of initial stresses is unlikely to be a realistic representation of poorly relaxed systems (e.g. strongly pre-sheared systems that may rather be expected to have a non symmetric distribution of initial stresses).


\section{From mesoscopic mean-field description to lambda-model}

We can actually derive from our model a dynamics of a specifically defined $\lambda$ fitting the general form of the lambda model \cite{wei2016quantitative,armstrong2016dynamic}. Just as  $\lambda$ does in the lambda model,  $s_d(t)$ (the standard deviation of the distribution at time $t$) can be seen as a phenomenological parameter characterizing the aging level in our model. However, as the model discusses the evolution of the full distribution $\mathcal{P}(\sigma,t)$, it is richer than a model that only uses one scalar parameter. Still, it is possible to make an approximate mapping of  $s_d(t)$ to $\lambda(t)$ by defining:

\begin{equation}\label{lambda}
\lambda\hat{=}e^{- \big(\frac{s_d}{ \sigma^{ext}}\big)^2}\;.
\end{equation}
As so, when $s_d$ gets small, $\lambda$ gets close to one corresponding to fully aged system, and when $s_d$ gets large, $\lambda$ gets close to zero corresponding to a rejuvenated system \cite{wei2016quantitative,armstrong2016dynamic}. 

The time evolution  of $\lambda$ is then  determined by that of $s_d^2=\langle\sigma^2\rangle-\langle\sigma\rangle^2=\langle\sigma^2\rangle-(\sigma^{ext})^2$. Since $\sigma^{ext}$ is fixed, $\frac{d}{dt}\big(s_d ^2\big)=\frac{d}{dt}\langle\sigma ^2\rangle$. By integrating on both sides of Eq(1) in the main text of paper with $\int d\sigma\sigma^2\bullet$, one obtains: 

\begin{equation}\label{sd2evolution}
\frac{d}{dt}\big(s_d^2\big) = 2G_0\sigma^{ext}\dot{\gamma}(t) - \frac{1}{\tau}\int_{|\sigma|>\sigma_c}d\sigma\mathcal{P}(\sigma,t)(\sigma^2 - 2\alpha)
\end{equation}
with

\begin{equation}\label{dotgamma}
\dot{\gamma}(t) = \frac{1}{\tau}\int _{|\sigma|>\sigma_c}d\sigma\sigma\mathcal{P}(\sigma,t)\;.
\end{equation}

Here for simplicity, we only consider a positive stress is applied, which insures $\dot{\gamma}(t)\geq 0$. Since $\sigma_c=1$ and $\alpha<0.5$, the integration in the second term of Eq.\ref{sd2evolution} is strictly positive. This also gains clear interpretation: the first term in Eq.\ref{sd2evolution} representing the flow contributes to rejuvenate the system, while the second term representing the local relaxations contribute to aging. By changing variables, one easily obtains the dynamics of $\lambda$: 

\begin{equation}\label{lambda_evolu}
\frac{d}{dt}\lambda = - \frac{2G_0}{\sigma^{ext}}\lambda\dot{\gamma} + \frac{1}{\tau \big(\sigma^{ext}\big)^2}\lambda K
\end{equation}
with $K(>0)$ is the integration in the second term of r.h.s. of Eq.\ref{sd2evolution}. $\dot{\gamma}$ and $K$ are in effect two functionals of $\mathcal{P}(\sigma,t)$, however they can be roughly regarded as two functions of $\lambda$ and $\sigma^{ext}$ if a simple parametrization (Gaussian or exponential) of $\mathcal{P}(\sigma,t)$  is chosen.

Comparing Eq.\ref{lambda_evolu} with the general form of lambda model \cite{wei2016quantitative}:

\begin{equation}\label{lambda-model}
\frac{d}{dt}\lambda = -f(\phi)\lambda^n + g(\phi)(1-\lambda)^m\;.
\end{equation}
Both formulations contain, on the r.h.s., a positive aging term and a negative term of rejuvenation driven by the external loading. It is not difficult to verify that the aging term in Eq.\ref{lambda_evolu} vanishes at $\lambda\rightarrow 1$ and the rejuvenation term vanishes at $\lambda\rightarrow 0$. Both the aging and rejuvenation terms are positive functions of $\sigma^{ext}$. Besides keeping the consistency with the lambda model, Eq.\ref{lambda_evolu} not only introduces more complexity on the interplay between the external loading and the structural parameter $\lambda$, but also offers a more generic physical picture of the widely studied lambda model from a mesoscopic point of view, i.e. the structural parameter can be related with mesoscopic stress fluctuations.

\section{Numerical Implementation}

We adopt the finite-difference Euler integration method to numerically solve the mean-field model described by Eq.(1) and the  schematic evolution described  by Eq.(4) and Eq.(5) in the main article.

We discretize the support of the time dependent probability density of local shear stresses $\mathcal{P}(\sigma,t)$ by $d\sigma$ within a domain of $\left[\sigma_\mathrm{min}, \sigma_\mathrm{max}\right]$ and discretize the time by $dt$. The total number of points in stress space is then $N=(\sigma_\mathrm{max}-\sigma_\mathrm{min})/d\sigma$. The probability density $\mathcal{P}(\sigma,t)$ is therefore discretized into $\mathcal{P}_i(t)=\mathcal{P}(\sigma_i,t)$, with $\sigma_i=\sigma_\mathrm{min}+i\times d\sigma$ and $i=0,1,...,N-1$. A periodic boundary condition is used, so that $P_N=P_0$. The source term Dirac function $\delta(\sigma)$ is approximated by a discretized Gaussian function with a narrow standard deviation, i.e. $\delta_i=\frac{1}{\xi _1\sqrt[]{2\pi}}\exp(-\sigma_i^2/\xi_1^2)$ and the step function $\theta(|\sigma|-\sigma_c)$ is approximated by $\theta_i = \left[\tanh(-\xi_2(\sigma_i+\sigma_c))+\tanh(\xi_2(\sigma_i-\sigma_c))+2\right]/2$. The two parameters $\xi_1$ and $\xi_2$ control respectively the width of the Gaussian and the width of the region with large variation of $\theta_i$. Taking the limits $\xi_1\rightarrow0$ and $\xi_2\rightarrow\infty$, $\delta_i$ and $\theta_i$ converge to $\delta(\sigma)$ and $\theta(|\sigma|-\sigma_c)$ respectively, along which we should also take $d\sigma\rightarrow0$ for the numerical discretization reflects well the regularity of these functions, whereas the trade off being the computing cost.

We use a symmetrised finite difference method for estimating the first and second partial derivatives with respect to stress, as follows:
\begin{eqnarray}\label{1std}
\partial_{\sigma}\mathcal{P}\big|_i(t)= \frac{\mathcal{P}_{i+1}(t)-\mathcal{P}_{i-1}(t)}{2d\sigma}.
\label{eq:1std}
\end{eqnarray}
\begin{eqnarray}\label{2std}
\partial_{\sigma}^2\mathcal{P}\big|_i(t)= \frac{\mathcal{P}_{i+1}(t)-2\mathcal{P}_i(t)+\mathcal{P}_{i-1}(t)}{d\sigma^2}.
\label{eq:2std}
\end{eqnarray}
Periodic boundary conditions are applied for estimating the partial derivatives at the two edges. We therefore compute at each time step $t$ with the current probability density $\mathcal{P}_i(t)$ all necessary quantities for performing a Euler integration for the time evolution, i.e.
\begin{eqnarray}\label{evol}
\mathcal{P}_i(t+dt) &=& \mathcal{P}_i(t)+dt \bigg(-G_0\dot{\gamma}(t)\partial_{\sigma}\mathcal{P}\big|_i(t) \nonumber\\
&& + \alpha\Gamma(t)\partial_{\sigma}^2\mathcal{P}\big|_i(t)  -\frac{1}{\tau}\theta_i\mathcal{P}_i(t) + \Gamma(t)\delta_i \bigg)\;.
\label{eq:evol}
\end{eqnarray}
where $\Gamma(t)=\frac{1}{\tau}\sum_id\sigma\theta_i\mathcal{P}_i(t)$. For implementing the shear rate control protocol, $\dot{\gamma}(t)$ is simply set to a constant equal to the desired value. For implementing the stress control protocol, we compute $\dot{\gamma}(t)=\frac{1}{G_0\tau}\sum_id\sigma\theta_i\sigma_i\mathcal{P}_i(t)$.

We compared the solution of a pure diffusion equation solved by our numerical method and the analytical solution with initial conditions as a narrow Gaussian centered at zero. Generally the condition $dt\lesssim d\sigma^2/2$ is required for the numerical solutions to be stable and to approximate well the true solution for a diffusion equation. As long as the width of the probability density is much less than the numerical domain $[\sigma_\mathrm{min},\sigma_\mathrm{max}]$, our numerical solution compares well with the analytical one (data not shown here).  

From the analytical stationary solution of the H\'ebraud-Lequeux model \cite{agoritsasEPJE15,EliArxiv16}
, we know that, for the stress range that we are interested for studying the creep behavior, the probability density $\mathcal{P}(\sigma)$ is mainly weighted between $0$ and $\sigma_c=1$ and the stationary solution beyond an absolute stress value of two $\mathcal{P}_{stat}(|\sigma|>2)$ is quasi-null. Thus we restrict the numerical domain constraint by $\sigma_\mathrm{min}=-5$ and $\sigma_\mathrm{max}=5$, which is justified by our numerical solutions with $\mathcal{P}_{i=0}\approx\mathcal{P}_{i=N-1} \lesssim 10^{-8}$ during its evolution. As mentioned before, the stress discretization $d\sigma$ is determined by how much we want $\theta_i$ and $\delta_i$ to be, respectively close to the real step function and the Dirac function, so that we try to take as small as possible the numerical domain to reduce the computational cost with respect to the number of points in stress space. 
\begin{figure}[th]
	\begin{center}
		\includegraphics[width=\columnwidth, clip]{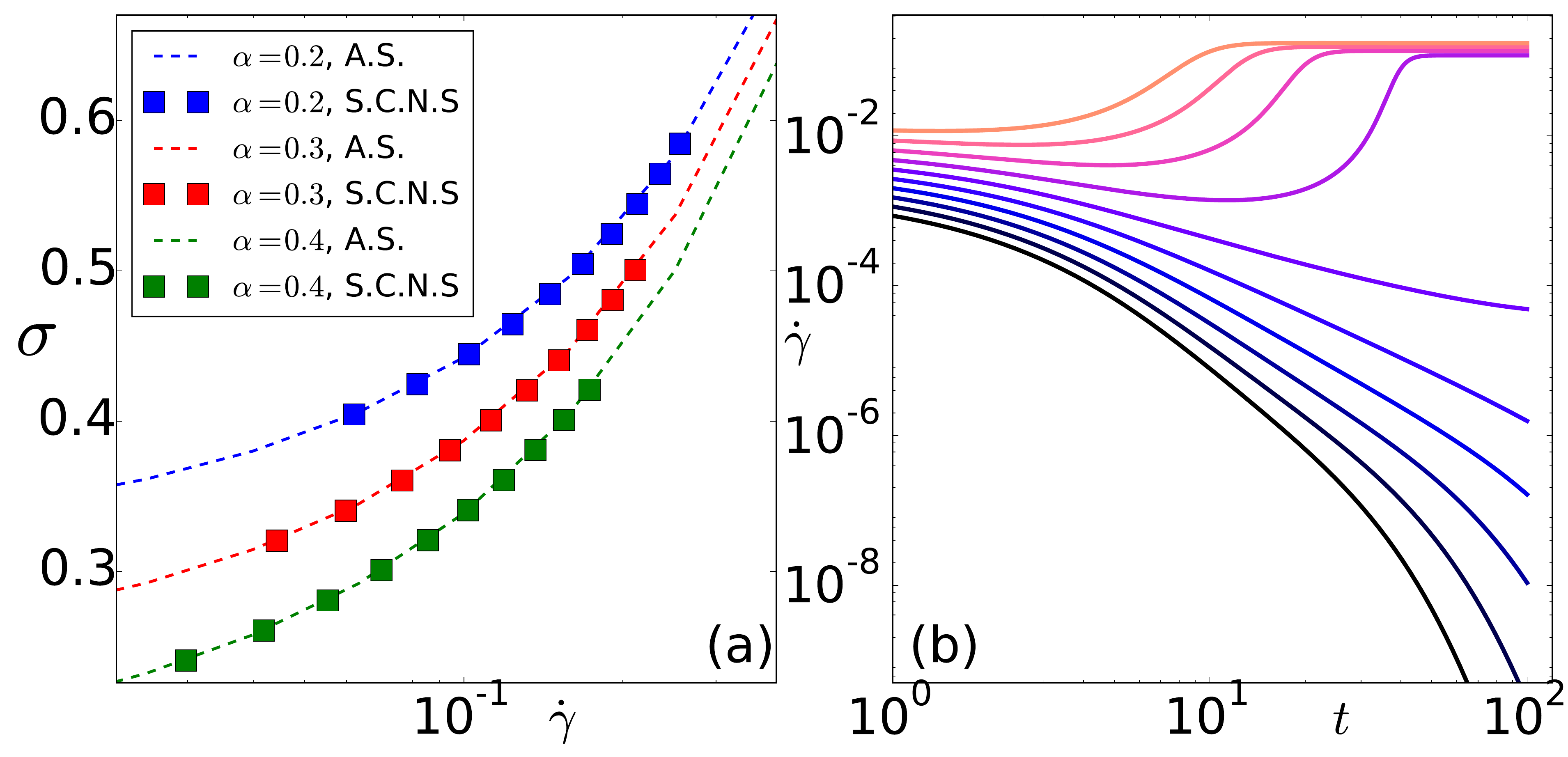}
	\end{center}
	\caption{{} (a)Flow curves for different values of $\alpha$ with different colors. Dashed lines: the analytical solution; Squares: the flow curves obtained by our creep simulations within steady state regime. (b) Creep curves $\dot{\gamma}(t)$ for $(\alpha=0.4, s_d=0.26)$ and from the bottom to the top $\sigma^{ext}-\sigma_y=0.1, 0.12, ..., 0.28$, the extrapolated static yield stress $\sigma^S_y(\alpha=0.4, s_d=0.26)\approx 0.185 + \sigma_y$. All curves with $\sigma<\sigma^S_y$ seem to vanish exponentially in time.}
	\label{fig:1}
\end{figure}

Our discretization $d\sigma$ is chosen according to the  following considerations. Firstly, the spacial precision around $\sigma=\pm \sigma_c = \pm 1$ is crucial for well solving the dynamics for some parameter settings when $s_d$ and $\sigma^{ext}$ are both small, because a very small part of the probability density is present beyond $\pm\sigma_c$. It has been shown that the dynamics of this boundary layer is crucially determining the overall dynamics \cite{EliArxiv16}. Another constraint is coming from the regularization of the derivatives of the Heaviside function. The width of the approximated function must be small enough with respect to the width of the boundary layer (given by $\mathcal{P}(|\sigma|> \sigma_c)$) to approximate well the dynamics. Thus the choice of $d\sigma$ needs to be small enough to both well discretize the regular form of $\mathcal{P}(\sigma)$ and the width of the approximated step function $\theta_i$, otherwise some uncontrollable errors may occur. But note that decreasing $d\sigma$ increases the computing cost within one time step. Besides, since $dt\lesssim d\sigma^2/2$ is required, fluidization time being fixed, decreasing $d\sigma$ also means increasing the number  of computing time steps for reaching the stationary state during one creep simulation. Taking all the above constraints into account, we have to tune $d\sigma$ and $dt$ to compromise between the computing cost and the domain of parameter settings $(\sigma^{ext},s_d)$ that can be faithfully explored by our code. For obtaining the results in the main article, we set $dt=10^{-5}$ and $d\sigma = (\sigma_\mathrm{max}-\sigma_\mathrm{min})/8192 \approx 10^{-3}$, which is at the limit of the stability condition $dt\lesssim d\sigma^2/2$. The results coming from this numerical discretization are compared with those coming from a numerical discretization where $dt$ fully satisfying the stability condition, and only a tiny difference is observed which does not affect quantitatively the physics of creep. As a benchmark of our numerical method, we compare our numerical steady state flow curves with those obtained by implicit analytical stationary solutions of the H\'ebraud-Lequeux model \cite{agoritsasEPJE15,EliArxiv16}. The good agreement shown in Fig.\ref{fig:1}(a) validates our simulations. 

With the above discretization settings and within the acceptable simulation duration, for some parameter settings with both small values of $s_d$ and $\sigma^{ext}$, we still did not observe the fluidization regime in the creep curve $\dot{\gamma}(t)$ and the shear rate $\dot{\gamma}$ goes down with time until the end of simulation, as example in Fig.\ref{fig:1}(b). With our hypothetical relation $\tau_f=A(s_d, \alpha)\big[\sigma-\sigma^S_y(s_d, \alpha)\big]^{-\beta_s(s_d, \alpha)}$ justified by the figure(2.c) in the main article, we extrapolate the $\tau_f(s_d, \sigma; \alpha)$ for these parameter settings if $\sigma>\sigma^S_y(s_d,\alpha)$. If $\sigma<\sigma^S_y(s_d,\alpha)$,  $\tau_f=\infty$ meaning that at some point $\dot{\gamma}(t)$ vanishes exponentially. It is found out that in the first case when $\sigma>\sigma^S_y(s_d,\alpha)$, the $\tau_f$ extrapolated for these parameters are out of the scope of our total simulation time and in the second case $\sigma<\sigma^S_y(s_d,\alpha)$, $\dot{\gamma}(t)$ decreases faster than a power law, as the bottom curves in Fig.~\ref{fig:1}(b). These observations are not contradictory with but even support our interpretations. 


\bibliography{biblio_creep}

\end{document}